\documentclass[mathleft,fleqn,%
]{an}
\usepackage{graphicx}
\usepackage[varg]{txfonts}
\usepackage{natbib}
\usepackage{url}
\usepackage{xspace}

\newcommand{\pmn}{PMN\,J1603$-$4904\xspace}

\bibpunct{(}{)}{;}{a}{}{,}
\begin{document}

% The following seven commands are intended for editorial usage and
% should be ignored by the author(s).
\Pagespan{1}{}% Document's page range. 
% If second parameter is left empty, the last page is computed
% automatically.
\Yearpublication{2016}%
\Yearsubmission{2016}%
\Month{0}%   
\Volume{999}%  
\Issue{0}% 
\DOI{asna.201400000}% 

\title{Multiwavelength and parsec-scale properties of extragalactic Jets}
\subtitle{Doctoral Thesis Award Lecture 2015}

\author{Cornelia M\"uller\inst{1,2,3}\fnmsep\thanks{Corresponding author:
        {c.mueller@astro.ru.nl}}
}
%\titlerunning{Multiwavelength and parsec-scale properties of southern
%extragalactic Jets}
\authorrunning{C.\,M\"uller}
\institute{
Department of Astrophysics/IMAPP, Radboud University Nijmegen, 6500
GL, Nijmegen, The Netherlands 
\and 
Dr. Karl-Remeis-Sternwarte and ECAP, 96049 Bamberg, Germany
\and 
Lehrstuhl f\"ur Astronomie, Universit\"at W\"urzburg, 97074 W\"urzburg, Germany
}
\received{XXXX}
\accepted{XXXX}
\publonline{XXXX}


\abstract{Extragalactic jets originating from the
central supermassive black holes of active galaxies are powerful, highly relativistic plasma outflows, emitting light from the radio up to the
$\gamma$-ray regime. The details of their formation, composition and emission mechanisms
are still not completely clear. The combination of high-resolution
observations using very long baseline interferometry (VLBI) and
multiwavelength monitoring provides the best insight into these
objects. Here, such a combined study of sources of the TANAMI sample is presented, investigating the parsec-scale and
high-energy properties. The TANAMI program is a
multiwavelength monitoring program of a sample of the radio and $\gamma$-ray brightest extragalactic jets in the
Southern sky, below $-30^\circ$\,declination. We obtain the
first-ever VLBI images for most of the sources, providing crucial information on the jet kinematics and
brightness distribution at milliarcsecond
resolution. Two particular sources are discussed in detail: \pmn, which can be
classified either as an atypical blazar or a $\gamma$-ray loud (young)
radio galaxy, and Centaurus~A, the nearest radio-loud active
galaxy. The VLBI kinematics of the innermost parsec of Centaurus~A's jet
result in a consistent picture of an accelerated jet flow with a
spine-sheath like structure. 
}
\maketitle

%%%%%%%%%%%%%%%%%%
\section{Introduction}\label{sec:intro}
Active galactic nuclei (AGN) emit light across the whole
electromagnetic spectrum, often dominating the emission of their host
galaxy.  Due to accretion onto the central supermassive black hole
(SMBH), they can produce so-called ``jets'', highly relativistic plasma
outflows. They belong to the most fascinating objects in the Universe,
but the underlying physics is still not
fully understood. The knowledge is crucial in context of AGN
feedback and multimessenger astronomy.  Multiwavelength observations
are useful tools to address open questions concerning the formation,
acceleration and the mechanism(s) behind the broadband emission up to the
highest energies.  

Blazars are a subset of radio-loud AGN, where the jet is observed at a
small angle to the line of sight, such that the jet emission is strongly
Doppler boosted. They belong to the most luminous and highly variable
sources \citep{Urry1996variability}, typically showing superluminal motion in the pc-scale
radio jet \citep[e.g.,][]{Lister2013}.
With the detection of $\gamma$-ray emission of AGN jets by
\textsl{EGRET} \citep{Hartman1999} 
various models were considered in order to explain the broadband
emission \citep[e.g.,][and many more]{Marscher1985,Mannheim1993,Sikora1994,Dermer1997,Dermer2012,Boettcher2013}.
A typical radio to $\gamma$-ray SED of a blazar shows a double-humped spectral shape from the
radio up to the $\gamma$-ray regime \citep{Fossati1998}.  While the low-energy peak can be well
explained by synchrotron emission, it is still discussed which
emission processes are responsible for the high-energy peak. It is
contentious whether it is due to synchrotron self-Compton
up-scattering and/or inverse Compton
scattering with external photons. Furthermore, the composition of the
ejected plasma, leptons or hadrons or the combination of both, plays
an important role in modeling the broadband emission.
Single-zone leptonic models have been very successful in describing
the broadband spectrum, however they fail to explain observations revealing rapid flaring and
multiple emission zones. In that case models need to take the jet
geometry into account \citep[like spine-sheath configuration,
e.g.,][]{Tavecchio2008}. Hadronic models, on the other hand, attempt
to explain the high-energy hump due to accelerated hadrons inducing
pion-photo production resulting in a electromagnetic cascade.

The combination of simultaneous broadband data allows us to study the
spectral energy distribution (SED) and the variability across the
bands. This provides information on the different radiating
components, e.g., the disk, broad line region or the jet, all together
making up the overall spectrum. Since these sources show strong
variability across all wavelengths, simultaneity of the data is
essential, i.e., contemporaneous monitoring at different wavelengths
is required.

In addition to the broadband spectral data, we use high-resolution
radio data from Very Long Baseline Interferometry (VLBI). It is a
unique tool to address the innermost regions of extragalactic jets at
milliarcsecond (mas) scales. It provides the highest angular
resolution and insights into regions close to the jet base where the
high-energy emission is thought to be produced. VLBI images reveal the morphology
of the jets at (sub-)pc scales. Typical blazar jet morphologies
are compact or one-sided, while for
larger jet inclination angles, where relativistic beaming effects are small, the jet and the counterjet can be
detected \citep[see, e.g.,][]{Kadler2004}.

Most objects detected with the Large Area Telescope (LAT) onboard of the \textsl{Fermi} Gamma-ray Space Telescope are classified as blazars
\citep{2fgl,3fgl}. Only few of the so-called ``misaligned'' objects (radio galaxies) with 
jets seen edge-on
\citep{Abdo2010_misaligned} are bright in the $\gamma$-rays. However,
these objects are of particular interest and challenge theoretical jet emission
models, which typically explain the high-energy spectral component with
high beaming factors. Their study can help to determine the
$\gamma$-ray emission region(s) and to constrain emission models,
because the broadband emission is less dominated by the beamed jet
emission \citep{Abdo2010_cenacore}.
In the radio regime, these misaligned objects can be
divided into evolved (e.g., as Centaurus~A or M\,87) and young radio
galaxies. The jets of the former have sizes up to several hundred
kiloparsecs, while the latter are typically more compact and smaller than
1\,kpc. Therefore, these sources are also called Compact
Symmetric Objects \citep[CSO,][]{ODea1998,Readhead1996a,Readhead1996b}. Because of
their intrinsic power, theoretical models predicted $\gamma$-ray emission from CSOs
\citep{Stawarz2008,Kino2007,Kino2009}, but no detection is
confirmed yet.

Here, the multiwavelength and VLBI study of extragalactic jets on
the Southern Hemisphere is presented. This work was performed in the
framework of the multiwavelength monitoring program TANAMI
(Sect.~\ref{sec:tanami}). After a short introduction to the project
and the sample results (Sect.~\ref{sec:tanamisources}), the properties
of two particular
sources are discussed, \pmn (Sect.~\ref{sec:pmn}) and Centaurus~A  (Sect.~\ref{sec:cena}).

%%%%%%%%%%%%%%%%%%
\section{TANAMI - The multiwavelength monitoring program of
  extragalactic jets in the Southern Hemisphere}\label{sec:tanami}
% incl. data reduction
In 2007, the TANAMI\footnote{\textit{Tracking Active Galactic Nuclei with
Austral Milliarcsecond Interferometry}
\url{http://pulsar.sternwarte.uni-erlangen.de/tanami/}} program
started monitoring the brighest extragalactic jets in the Southern
sky (below $-30^\circ$ declination) using the combination of
high-resolution VLBI observations at 8.4\,and 22.3\,GHz \citep{Ojha2010a}
and corresponding radio
monitoring with the Australian Compact Array
\citep[ATCA,][]{Stevens2012} and the Ceduna telescope \citep{Blanchard2012a}, in the optical/UV with
\textsl{Swift}/UVOT and \textsl{Rapid
  Eye Mount} \citep{Nesci2013}, in the X-rays with \textsl{Swift}/XRT and all-sky observations at $\gamma$-rays by
\textsl{Fermi}/LAT. 

In addition, pointed observations for particular
sources are conducted with \textsl{XMM}-Newton, \textsl{Suzaku}
\citep[see Sect.~\ref{sec:pmn} and e.g.,][]{Mueller2015a,Kreikenbohm2016} and
\textsl{INTEGRAL} in the X-rays.

\begin{figure}
\includegraphics[width=\columnwidth]{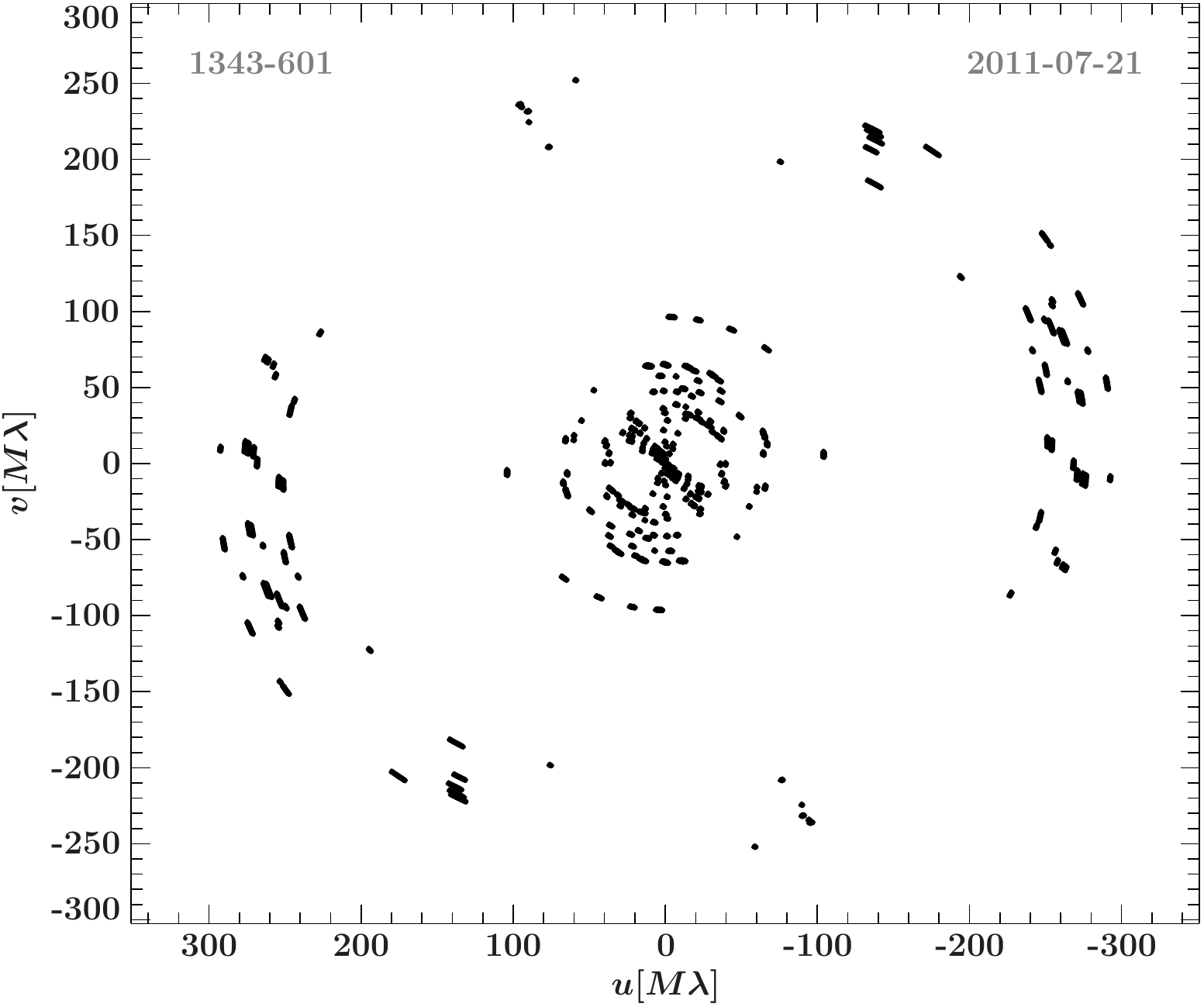}
\caption{The typical TANAMI $(u,v)$-coverage at 8\,GHz has improved due to the
  addition of new
  telescopes since 2011 compared to the initial array configuration
  presented in \citet{Ojha2010a}. Here, the
  intermediate $(u,v)$-range is covered due to baselines to Warkworth. The
  long baselines are provided by TIGO and Hartebeesthoek. The
  displayed $(u,v)$-coverage corresponds to the observation of
  Centaurus~B (PKS\,1343--601), shown in Fig.~\ref{fig:new}.}
\label{fig:uvplot}
\end{figure}

The TANAMI VLBI observations are performed with the Australian Long
Baseline Array (LBA), including additional telescopes at NASA's Deep
Space Network (DSN) located at Tidbinbilla, the South-African
Hartebeeshoeck antenna, the German Antarctic Receiving Station (GARS)
in O'Higgins (Antarctica), and the Transportable Integrated Geodetic
Observatory (TIGO) in Chile \citep{Ojha2010a}. Since 2011, the
$(u,v)$-coverage (see as an example Fig.~\ref{fig:uvplot}) at
intermediate baselines is significantly improved by the Warkworth (New
Zealand), Katherine and Yarragadee (Australia) antennas \citep{Kadler2015}.  TANAMI
observations typically have an angular resolution of about a few mas,
down to less than 1 mas, with the largest baselines to the
transoceanic antennas Hartebeeshoeck, TIGO and O'Higgins.

The VLBI data are recorded on the LBADRs (Long Baseline Array Disk
Recorders) and correlated on the DiFX software correlator at Curtin
University in Perth, Western Australia \citep{Deller2007}. The
correlated data are inspected, edited and fringe fitted 
in AIPS \citep[National Radio Astronomy Observatory’s
Astronomical Image Processing System software;][]{Greisen2003_AIPS}. The
amplitude calibration is performed using known flux values of prior
observed sources. Hybrid imaging and subsequent model fitting is performed
in the program DIFMAP \citep{Shepherd1997}.
For more details on the data reduction see \citet{Ojha2010a}, where we presented
the first-epoch images of the intial sample.

The TANAMI sample of extragalactic jets was defined as a hybrid radio
and $\gamma$-ray selected sample. Initially, it consisted of 43
objects, while to date $\sim$90 sources are regularly monitored. 
We aim to investigate the emission and formation mechanism of AGN jets.
Our observational setup provides structural and spectral information
at mas-scales, and, in addition, broadband spectral
properties with time \citep{Krauss2016}. 
The long-term VLBI monitoring yields information on jet properties, 
such as apparent speed, inclination, opening angles, and structural changes
with time. Furthermore, the dual-frequency approach provides spatial spectral
index distributions of individual jet features. 
The simultaneous broadband observations address the jet
activity and spectral changes across all wavelengths.

\begin{figure*}
\includegraphics[width=\columnwidth]{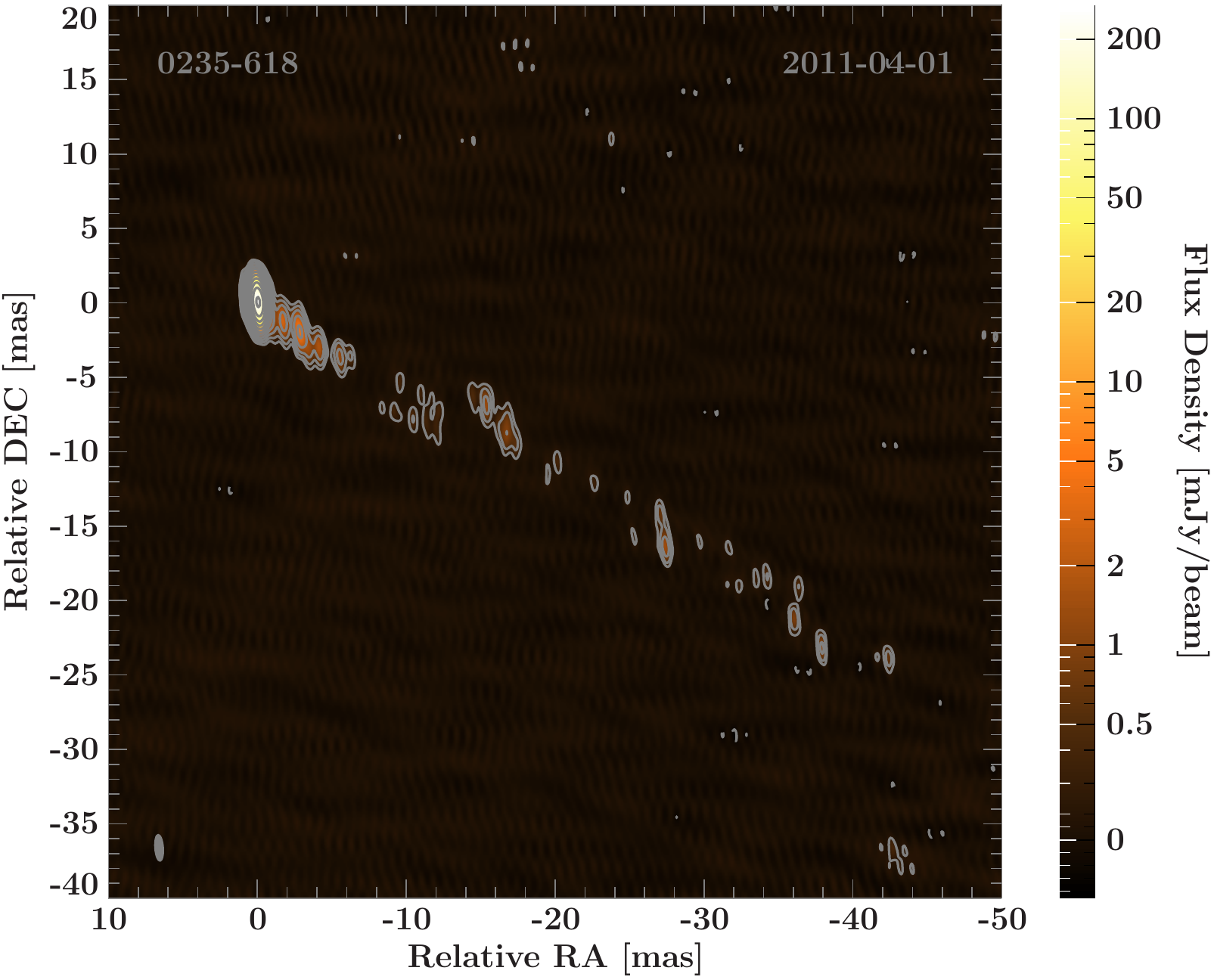}\hfill
\includegraphics[width=\columnwidth]{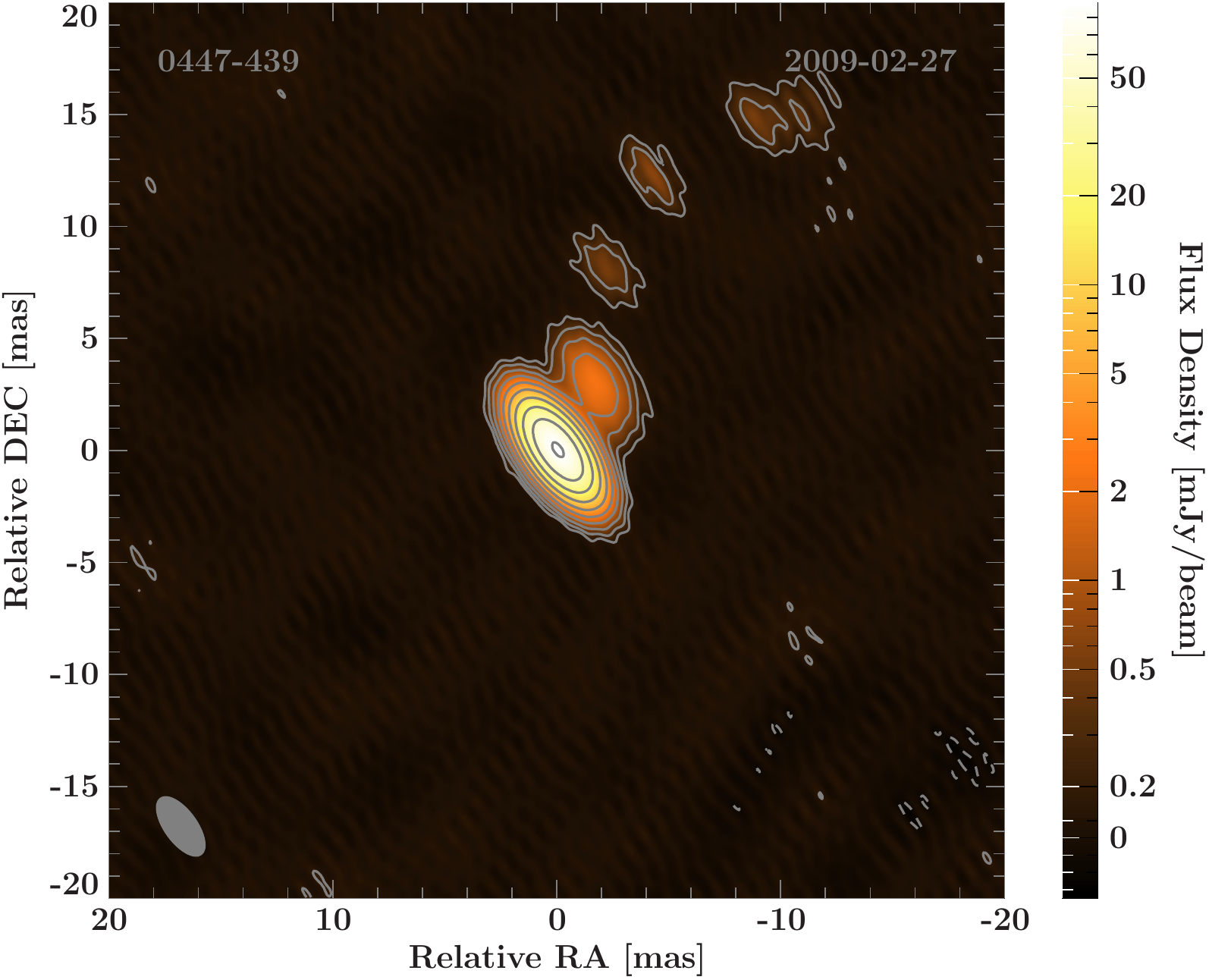}\hfill
\includegraphics[width=\columnwidth]{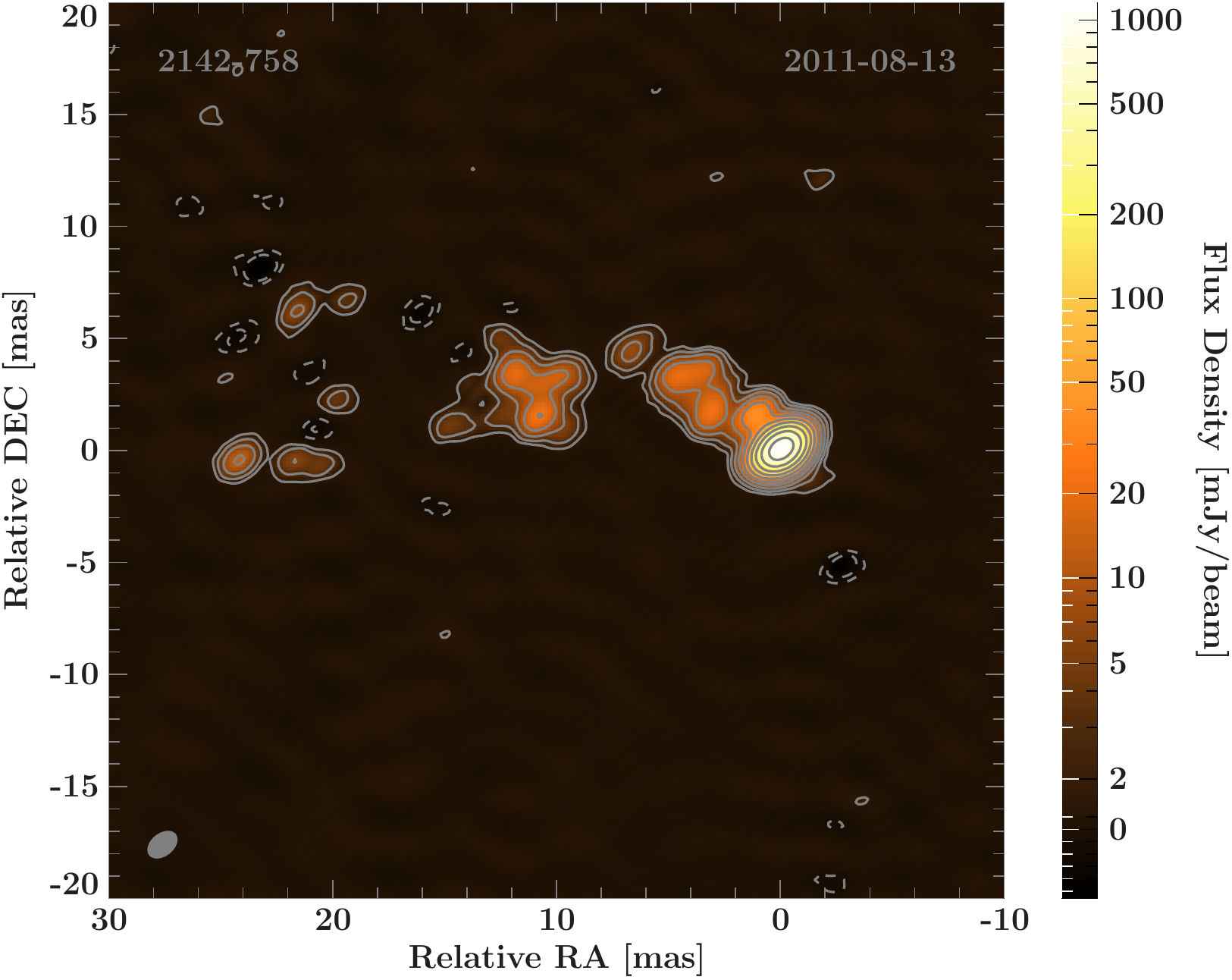}\hfill
\includegraphics[width=\columnwidth]{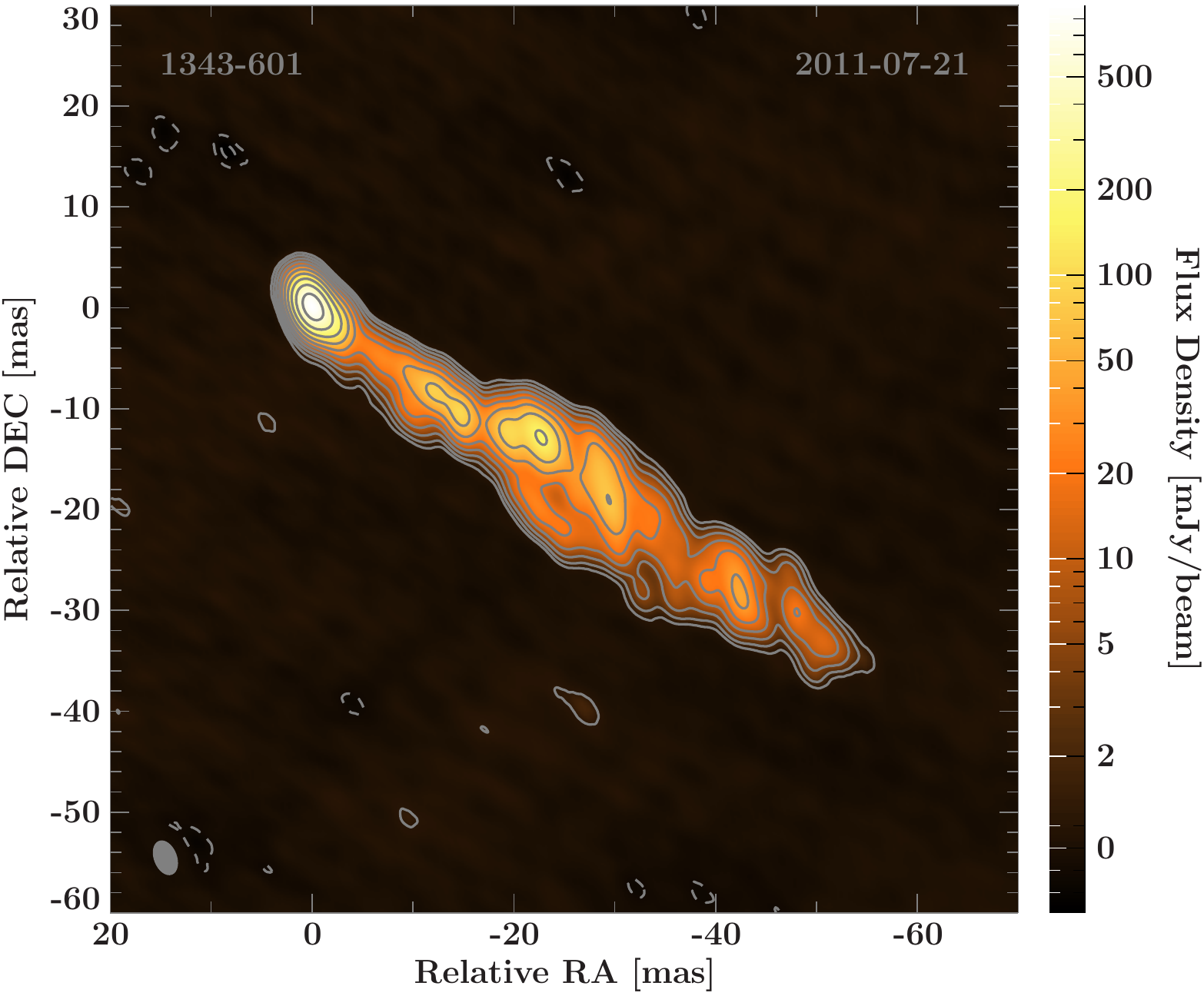}
\caption{First-epoch VLBI images from newly added TANAMI sources
  (M\"uller et al., in prep.). \textit{From top left to bottom right}:
  Parsec-scale morphology of 
  \mbox{PKS\,0235--618}, one of the blazar sources consistent with one
  of the PeV neutrino events detected by IceCube \citep[see][for more
  details]{Krauss2014a}, the TeV-blazar PKS\,0447--439, the flat-spectrum $\gamma$-ray quasar  PKS\,2142-758
  \citep[see][for details on the multiwavelength
  properties]{Dutka2013}, and the parsec-scale jet of
  the radio galaxy
  Centaurus~B (PKS\,1343--601) also detected at $\gamma$-rays
  \citep{Katsuta2013}. The color scale displays the flux density
  distribution in mJy/beam. Contours are scaled logarithmically,
  increased by a factor of 2, with the lowest contour set to the
  $3\sigma$-noise-level (negative contours are dashed). The restoring
  beam is shown as gray ellipse in the lower left corner. }
\label{fig:new}
\end{figure*}

Since the start of TANAMI, new flat-spectrum radio sources have been
added to the sample when associated with a $\gamma$-ray
detection by \textsl{Fermi}/LAT. For most of these sources, TANAMI
provides the first high-resolution VLBI images (M\"uller et al., in
prep.). Figure~\ref{fig:new} shows a selection of these mas-scale
CLEAN images from
8\,GHz VLBI observations.

We are
particularly interested in the so called
``radio-to-gamma-connection'', i.e., the correlation of high-energy
emission with changes in pc-scale properties seen in jets. This is
linked to the open question of the production sites and mechanisms of
high-energy photons. Furthermore, the multiwavelength analysis enables
us to study the broadband emission, to test different emission models,
related to the composition of jets \citep[e.g.,][]{Dutka2013}. 

Beyond that, TANAMI has a strong multimessenger component, trying to
explain the high-energy neutrinos observed with IceCube and ANTARES
with the broadband emission from $\gamma$-ray loud TANAMI blazars. The details
can be found in \citet{Krauss2014a},
\citet{AntaresTANAMI2015}, and \citet{Kadler2016}, as well as in \citet{FritschPhDT}, \citet{FehnPhDT} and
\citet{Mueller2014PhDT}, but it is
beyond the scope of this paper.

%%%%%%%%%%%%%%%%%%
\section{Gamma-ray loudness and milliarcsecond-scale properties of
TANAMI sources}\label{sec:tanamisources}
Using one year of contemporaneous TANAMI VLBI and \textsl{Fermi}/LAT
data, we addressed the high-energy properties and
radio-$\gamma$-correlation in our sample \citep{Boeck2016}.  We
analyzed the radio and $0.1-100$\,GeV $\gamma$-ray data obtained during
the first 11\,months of \textsl{Fermi}/LAT monitoring.  More
than 70\% of the TANAMI sources are associated with $\gamma$-ray
emitters detected by \textsl{Fermi}/LAT. Upper limits on the
$\gamma$-ray flux were determined for the remaining sources, yielding
three new significant detections, which were later confirmed by the
\textsl{Fermi}/LAT team \citep{2fgl}.

We find increasing brightness temperatures $T_B$ of the radio cores, obtained
from VLBI measurements, with 1\,year average $\gamma$-ray
luminosity. The $\gamma$-ray undetected sources have lower
$T_B$ values and probably
$\gamma$-ray luminosities close to the determined upper limits.

Comparing the morphologies of the $\gamma$-ray detected versus
undetected TANAMI sources, we find that the $\gamma$-ray loud objects
are generally more core dominated, i.e., are more compact and have a
higher core-to-jet flux ratios (M\"uller et al., in prep.).  
These results are consistent with the general picture of strong
Doppler boosting in $\gamma$-ray bright sources. 

Most $\gamma$-ray sources in our
sample show compact or one-sided jet morphologies. Two exceptional objects,
\pmn and Centaurus~A, are discussed in more
detail in the following sections. Their double-sided\footnote{Note
  that, as in \citet{Ojha2010a}, we use the morphology classification scheme by
  \citet{Kellermann1998}, classifying sources with the most compact
  component in the middle of the emission as double-sided, without taking
  spectral information into account.} morphology and
$\gamma$-ray loudness are in particular interesting and provide
important insights into the jet physics.

%%%%%%%%%%%%%%%%%%
\section{The unusual jet source \pmn}\label{sec:pmn}

The radio source \pmn has been known as a calibrator source (PKS\,1600$-$48)
for Southern Hemisphere radio observations, but it attracted attention due to
its association with one of the brightest, flat-spectrum $\gamma$-ray
sources detected during the first few months of \textsl{Fermi}/LAT
monitoring \citep{0fgl,Kovalev2009_LBAS}. Since then, it is
is a bright, significantly detected $\gamma$-ray source  \citep[3FGL\,1603.9--4903;][]{1fgl,2fgl,3fgl,1fhl,2fhl}.
Based on sparse optical and broadband data, the source was classified
as a low-synchrotron peaked BL\,Lac object without known redshift
\citep{Shaw2013a}. 

Due to the high-confidence association of the known radio source with the
$\gamma$-ray object 1FGL\,J1603.8$-$4903 \citep{1fgl}, it was included in the
TANAMI sample. The first VLBI observations were performed in 2009 and
revealed unusual pc-scale properties \citep{Mueller2014a}. This result
triggered
multiwavelength follow-up observations further questioning its classification as a blazar 
\citep{Mueller2015a,Goldoni2016}. 
In the following these observations and their conclusions will
be presented.

\begin{figure}
\includegraphics[width=\columnwidth]{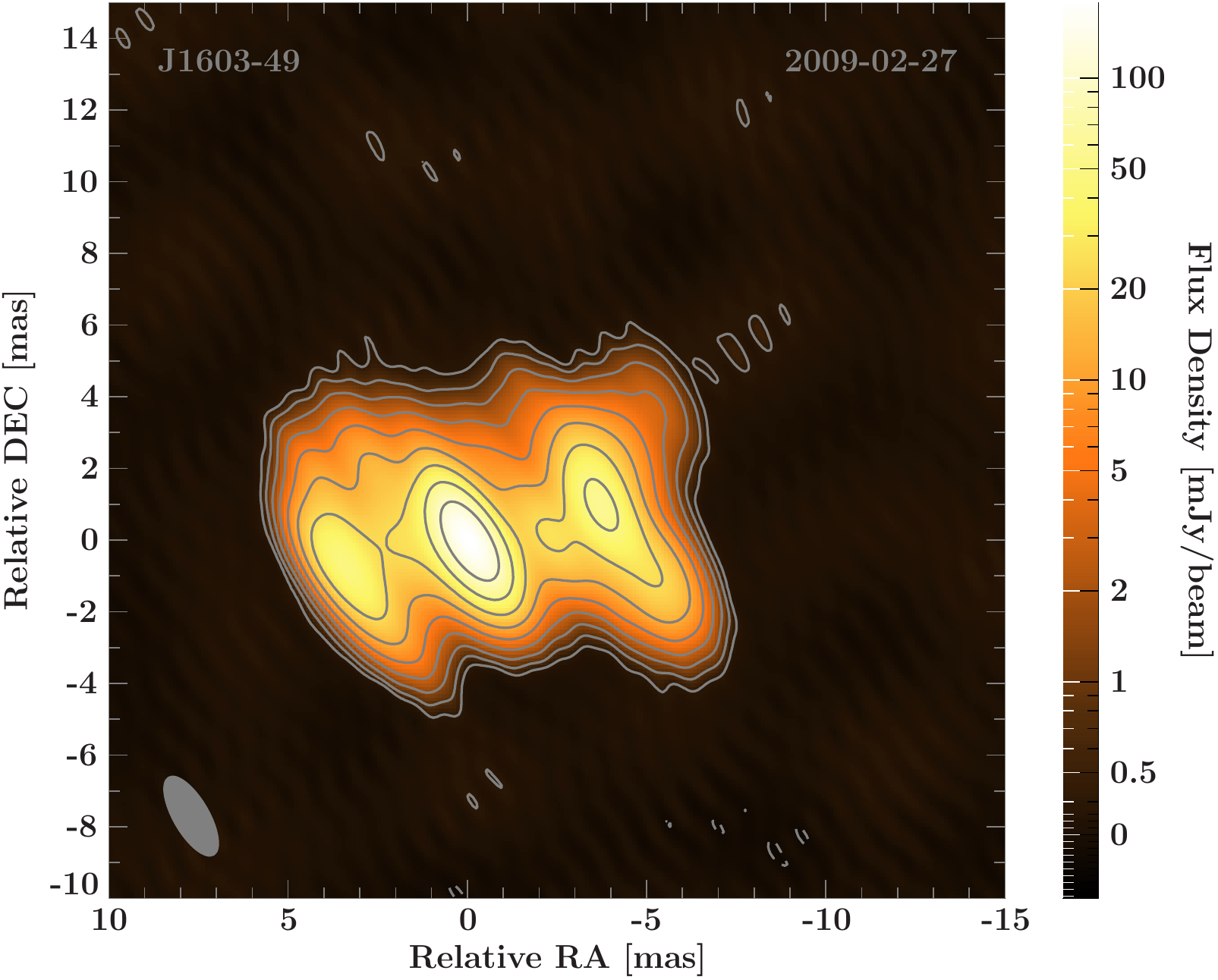}
\caption{Naturally weighted 8\,GHz VLBI image of \pmn. The brightest, most compact
  component is the central feature. The overall brightness
  distribution is almost symmetric and constant over a monitoring
  period of $\sim$4\,years. The color scale displays the flux density
  distribution in mJy/beam. Contour lines scale logarithmically and 
  increase by a factor of 2, with the lowest contour set to the
  $3\sigma$-noise-level. The restoring
  beam is shown as a gray ellipse in the lower left corner. }
\label{fig:pmn}
\end{figure}

\subsection{VLBI observations}
The TANAMI 8\,GHz VLBI observations of \pmn yielded the first
image at mas-scale resolution of its radio morphology \citep{Mueller2014a}. 
Since 2009 the source has been observed about twice a year at 8\,GHz,
with one (quasi-)simultaneous observation at 22\,GHz in 2010
May. Figure~\ref{fig:pmn} shows the mas-scale brightness distribution of the
source at 8\,GHz. It is almost symmetrical with a total correlated
flux density of $\sim 600$\,mJy. It is resolved and shows an East-West
orientation (position angle of $-80^\circ$) with three distinct emission
features. The brightest, most compact component is located at the
center, with a brightness temperature of $T_B\geq 1\times 10^{10}$\,K.
Using spectral index information from combined 8\,GHz and 22\,GHz
VLBI datasets, we concluded that this central component is the most
plausible `core' of \pmn.

TANAMI monitoring reveals no significant proper motion of the Eastern
and Western components. The source appearance as in
Fig.~\ref{fig:pmn} remains effectively constant. By modeling the source with three Gaussian
emission regions, we find relative motions of $v_\mathrm{app}< 0.2\,\mathrm{mas\,yr^{-1}}$ \citep{Mueller2014a,Mueller2014PhDT}, later
confirmed by \citet{Hekalo2015} using observations
spanning four years.

\subsection{Broadband spectrum and variability}
The ATCA monitoring between 1\,and 40\,GHz reveals only very minor flux
variability at higher frequencies. The ATCA spectral index of $\alpha=-0.4$
is consistent with the VLBI results for
the Eastern and Western features. However, ATCA measures about
$\sim$200\,mJy more flux as detected with VLBI, possibly indicating a
diffuse extended emission component 
which is resolved out by the TANAMI VLBI array. 

Continuous monitoring by \textsl{Fermi}/LAT shows no major flaring
activity as typically observed in blazar sources \citep{2fgl,3fgl}. \pmn has been
reported in both high-energy catalogs \citep[1FHL, 2FHL][]{1fhl,2fhl}, being a potential
candidate for detection in the TeV range by ground based
Cherenkov telescopes.
 
Using all available (including archival) multiwavelength data, we
constructed a non-simultaneous broadband SED (see Fig.~\ref{fig:sed}). It includes data from
TANAMI VLBI and ATCA in the radio, \textsl{2MASS} and \textsl{WISE}
in the infrared, \textsl{Swift}/UVOT, GMOS and NTT in the optical,
\textsl{Swift}/XRT, \textsl{XMM-Newton}, \textsl{Suzaku} in the X-rays and
\textsl{Fermi/LAT} in the $\gamma$-rays. We parametrize the broadband
spectrum with two logarithmic parabolas. In the
infrared, we see a strong excess, which can be modeled with a black
body spectrum.

\subsection{\textsl{XMM} and \textsl{Suzaku} results and follow-up
  VLT/X-shooter observations} 
\begin{figure}
\includegraphics[width=\columnwidth]{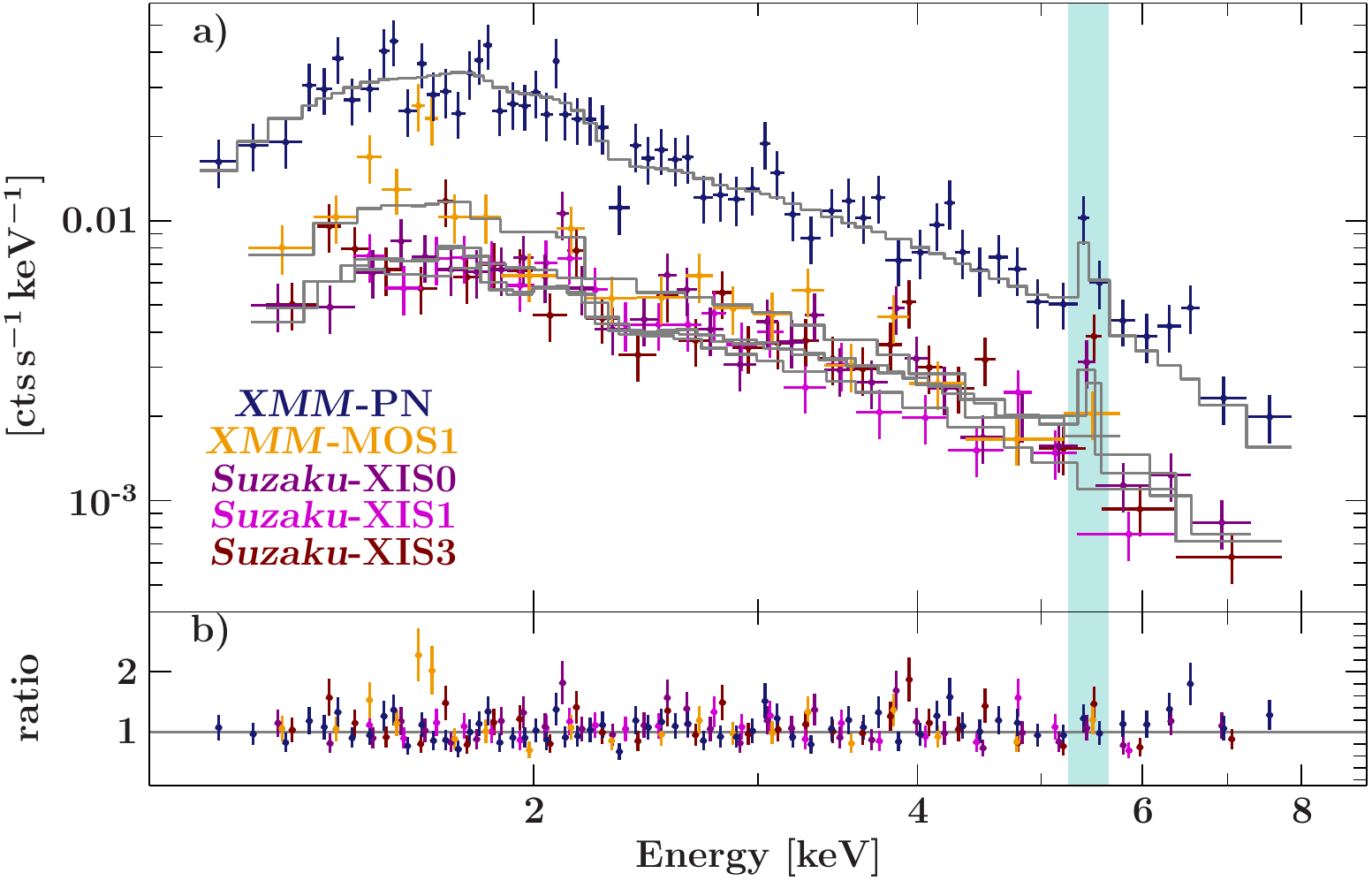}
\caption{Simultaneous fit to the (quasi-)simultaneous X-ray data from \textsl{XMM}-Newton and
  \textsl{Suzaku} of \pmn \citep{Mueller2015a}. The data are
  fitted by an absorbed power-law component and an emission line at
  $\sim$5.44\,keV (highlighted by the shaded region). a) Counts spectrum for all individual detectors. b)
  Ratio of data to model for the best fit the Gaussian
  emission line. The shaded region marks the position of the emission
  line. }
\label{fig:pmnspec}
\end{figure}

\begin{figure}
\includegraphics[width=\columnwidth]{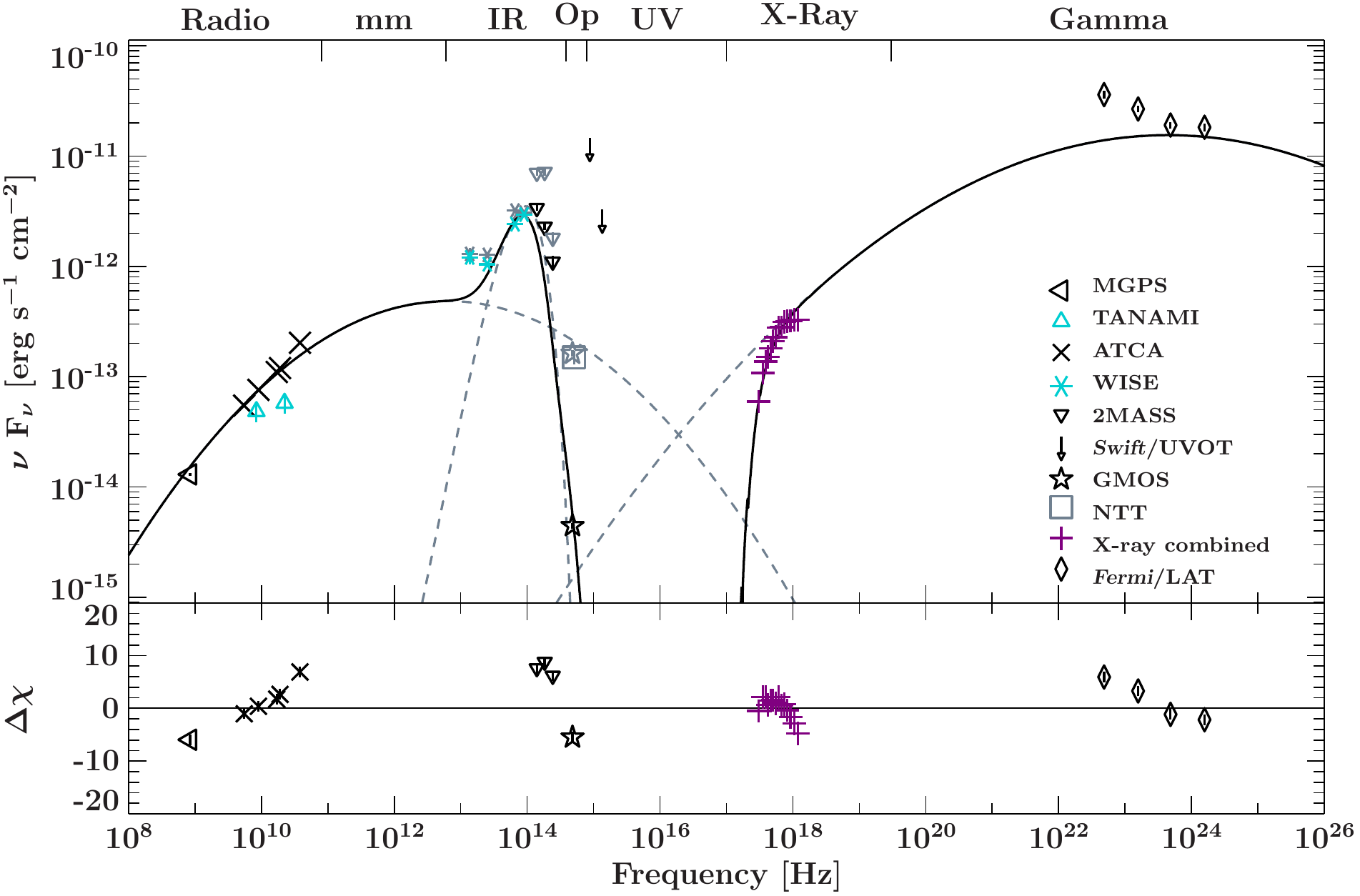}
\caption{Broadband $\nu F_\nu$ spectral energy distribution of \pmn,
  including VLBI total fluxes from TANAMI observations \citep{Mueller2014a}, data by
  MGPS \citep{Murphy2007_MGPS}, ATCA, WISE and 2MASS, by GMOS and \textsl{Swift}/UVOT 
  \citep{Mueller2014a} and NTT \citep{Shaw2013a}, and 4\,years data from
  \textsl{Fermi}/LAT from 3FGL \citep{3fgl}. The combined X-ray data
  by \textsl{Swift} XRT \citep{Mueller2014a}, \textsl{XMM}
  and \textsl{Suzaku} \citep{Mueller2015a} are shown in purple. We parametrize the data
  with two logarithmic parabolas (gray lines), absorbed by photoelectric absorption
  in the X-rays, and a blackbody component. The black symbols
  mark all data included in the fit, data marked by cyan symbols are excluded. The gray symbols represent
  the optical/IR and X-ray data corrected for extinction and
  absorption, respectively. The dashed lines show the unabsorbed log
  parabolas and the extinction corrected blackbody. In the lower panel
  we show the residuals of the fit in units of standard deviation of
  individual data points. A physical SED model is required to investigate
  the broadband emission in more detail.
}
\label{fig:sed}
\end{figure}

In order to better constrain the X-ray spectrum than with
\textsl{Swift}/XRT-only data, we performed \textsl{XMM}-Newton and \textsl{Suzaku}
observations in 2013 \citep{Mueller2015a}. The \mbox{2--10\,keV}
spectrum (Fig.~\ref{fig:pmnspec})
was simultaneously modeled with an absorbed power-law component
($N_\mathrm{H}=2.05_{-0.12}^{0.14}\times10^{22}\mathrm{cm^{-2}}$, $\Gamma=2.07_{-0.12}^{0.04}$),
cross-calibration constants for the different detectors, and
a Gaussian emission line at $5.44\pm0.05$\,keV. We interpret this X-ray line as the most
prominent emission line in AGN X-ray spectra, the 
Fe~K$\alpha$ fluorescence line with a restframe energy of
6.4\,keV. Therefore, the redshift of \pmn can be determined to be
$z=0.18\pm0.01$. Adopting this redshift, the brightness distribution
in the VLBI image (Fig.~\ref{fig:pmn}) has an extent of $\sim$46\,pc.

Triggered by these intriguing results, \citet{Goldoni2016} observed
\pmn with the UV-NIR VLT/X-shooter spectrograph in 2014. The optical continuum
spectrum is mostly featureless, not compatible with a stellar origin,
but likely non-thermal. Three emission lines are detected, allowing
us to determine a redshift of \mbox{$z=0.2321\pm0.0004$}. The spectral
features do not follow the definition of a BL\,Lac object. The new
redshift measurement implies that the X-ray emission line detected by
\textsl{XMM}-Newton and \textsl{Suzaku} should be interpreted as a
6.7\,keV line. This result is very peculiar, due to a missing
6.4\,keV line and is possibly indicating that
the 6.7\,keV emission is due to collisionally ionised plasma, and
merits further follow-up observations of \pmn.

\subsection{An unusual blazar or a $\gamma$-ray loud young radio
  galaxy?}
\pmn has been classified as a BL\,Lac object, i.e., a jet source
pointing towards the observer. The emission of these objects are
dominated by relativistic beaming effects and spectral features
like emission lines are usually outshined by the non-thermal jet
emission. Furthermore, blazars are typically highly variable across
the electromagnetic spectrum and VLBI measurements reveal compact,
sometimes one-sided jet structures, showing high polarization and
relativistic motion.

Compared to this, the VLBI and multiwavelength data of \pmn indicate very
unsual properties for a blazar source. 
Our VLBI data show a symmetric brightness distribution on mas-scales, which is
constant over 4\,years of monitoring. The brightest component in the
center has the flattest spectrum and highest brightness temperature
values. We find no significant proper motion.
Similarly, multiwavelength monitoring
reveals no rapid flaring activity. Only long-term variability is
reported at GeV energies \citep{1fhl,3fgl} and ATCA
monitoring over more than ten years shows some minor long timescale
flux density variations at higher radio frequencies.  ATCA data also
give upper limits of $<1.2$\% on the polarization \citep{Murphy2010_AT20G},
though higher polarization fractions at VLBI scales cannot be excluded
so far.  The broadband SED shows an unexpected strong excess in the
infrared and a significant X-ray spectral line and high intrinsic
absorption. These are all features which are not observed in typical
blazar spectra. Finally, the recent VLT/X-shooter data suggest a non-BL\,Lac
nature.

This peculiar appearance of \pmn is questioning its blazar
classification, hence, we consider alternative interpretations
\citep{Mueller2014a}, favouring a young radio galaxy seen at a
larger  angle to the line of sight.
Its VLBI structure resembles a double-sided morphology as seen in
FR\,I radio galaxies, but on smaller scales, i.e. a young
version. These compact symmetric objects typically show low
variability and polarization \citep{Peck2000} and VLBI observations
suggest that they are seen edge-on. 
Our multiwavelength data are in agreement with the young radio galaxy
scenario. Adopting the measured redshift
\citep{Mueller2015a,Goldoni2016} and assuming edge-on geometry, the
source size probed by VLBI is well below the canonical limit for young radio sources of
1\,kpc, while the ATCA data limits the total size to be less than 3\,kpc.

Further multiwavelength data will help to substantiate this
alternative classification. For example, VLBI monitoring can reveal
opposed apparent motion, as expected for a jet-counterjet
system. Low-frequency radio observations below 1\,GHz can give important information about
a potential spectral turnover, as expected for young radio sources. 

If
confirmed, \pmn will add to the class of misaligned $\gamma$-ray
bright sources. Moreover, the confirmed detection of a $\gamma$-ray
loud CSO would help to determine the location of the high-energy
emission using broadband SED models.

%%%%%%%%%%%%%%%%%%
\section{Zooming into the closest radio-loud AGN: Centaurus
  A}\label{sec:cena}
At a distance of only 3.8\,Mpc \citep{Harris2010}, the radio galaxy
Centaurus~A (Cen~A) is an ideal target to study the innermost region
of an AGN and jet physics at highest linear resolutions \citep[for a
detailed review on this source see,
e.g.,][]{Israel1998}. At this distance, an angular resolution of 
1\,milliarcsecond corrseponds to only $\sim$0.018\,pc. Cen~A is
detected from the radio up to TeV energies
\citep{Aharonian2009} and considered as a source candidate of
ultra-high energy cosmic rays \citep{Clay2010} and extragalactic
neutrinos \citep{Icecube2009}.

\begin{figure}
\includegraphics[width=\columnwidth]{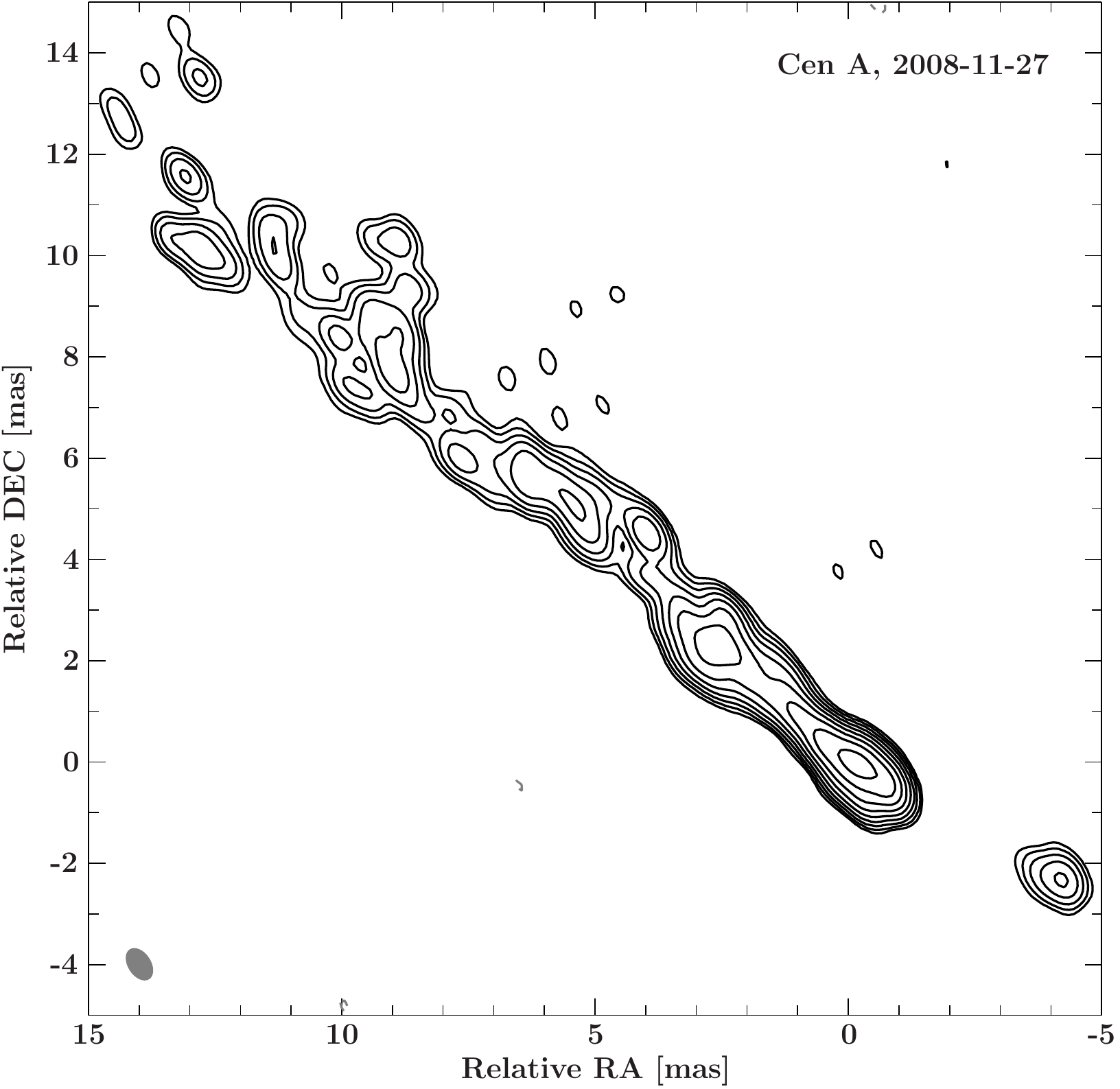}\\
\includegraphics[width=\columnwidth]{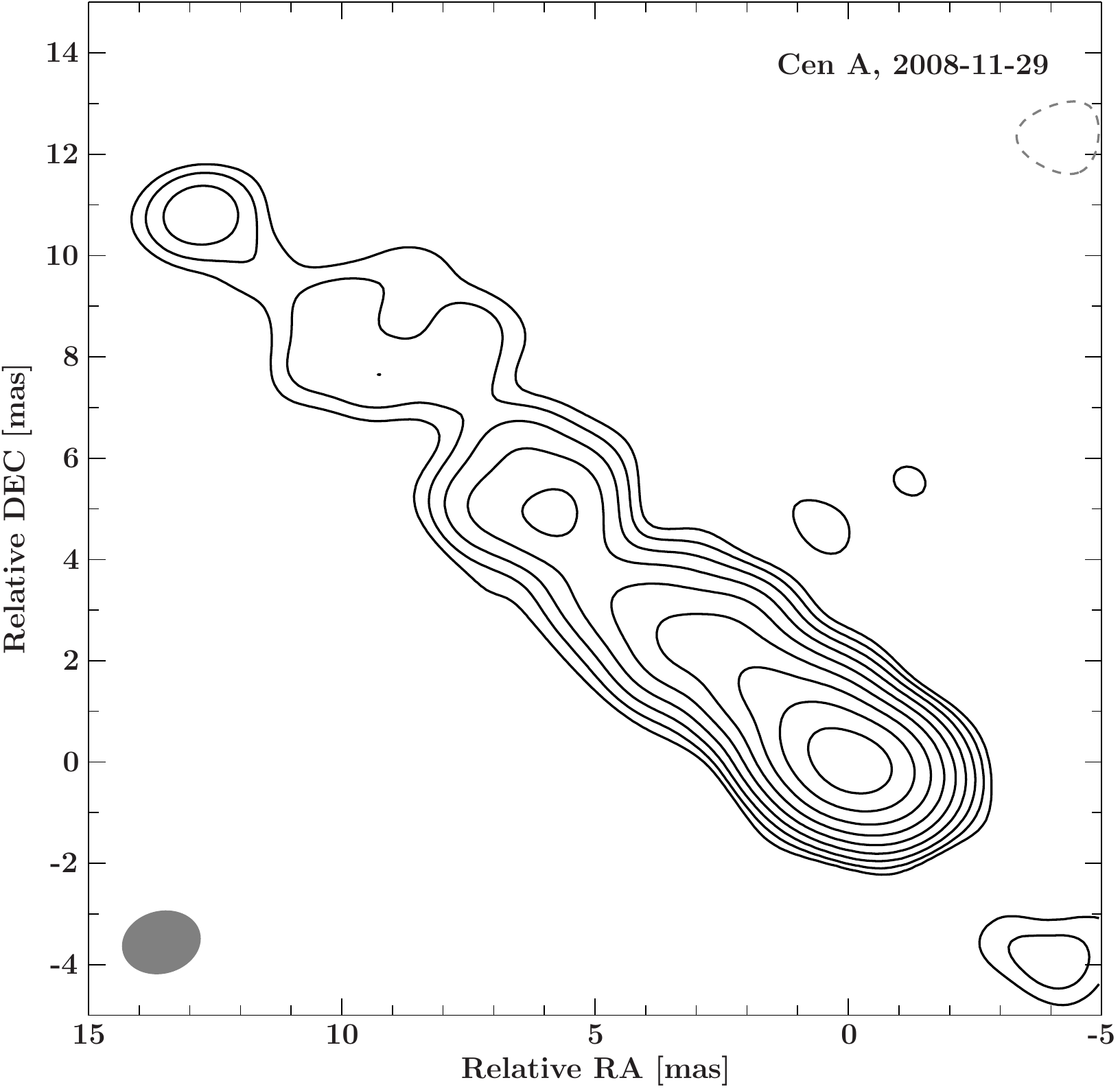}
\caption{Highest-resolution images of sub-pc scale morphology of
  Centaurus~A (top: 8.4\,GHz, bottom: 22.3\,GHz). Clean contour images
  from the dual-frequency observation in 2008 November 
 \citep{Mueller2011a} using uniform
  weighting. The ellipse in the lower
left corner of each panel indicates the restoring beam. 
}
\label{fig:cena2}
\end{figure}

In the framework of the TANAMI program Cen~A has been monitored about
twice a year with VLBI at 8\,GHz since 2007, including one
simultaneous epoch at 22\,GHz \citep{Mueller2014b,Mueller2011a}. Our
observations result in the highest resolved images of Cen~A (down to
$\sim$0.4\,mas corresponding to $\sim$0.007\,pc), allowing us to study
the jet-counterjet system in unprecedented
detail. Figure~\ref{fig:cena2} shows the imaging results for the
dual-frequency observation
\citep{Mueller2011a}. Figure~\ref{fig:epochs} shows the time evolution
of the first seven 8\,GHz images at natural weighting.
Table~\ref{tablecena} provides an overview of the observation logs for
these observations and corresponding image parameters.

Here, a summary of the results from this VLBI monitoring is presented,
based on the publications by \citet{Mueller2011a} and
\citet{Mueller2014b}.

\begin{table*}
\caption{Details on TANAMI observations of Centaurus~A and image parameters}
\footnotesize
\label{tablecena}
\begin{center}
\begin{tabular}{l c c c c c c r }
\hline\hline
Obs. Date \& Frequency & Array Configuration$^{a}$ & $S_\mathrm{peak}$$^{b}$ & RMS$^{c}$ & $S_\mathrm{total}$$^{d}$ & $b_\mathrm{maj}$$^{e}$ & $b_\mathrm{min}$$^{f}$ & P.A.$^{g}$ \\
(yyyy-mm-dd)  &                  & (Jy beam$^{-1}$) & (mJy beam$^{-1}$) & (Jy) & (mas) & (mas) & ($^\circ$) \\
\hline
\textbf{natural weighting} & & & & & & & \\
2007-11-10 (8.4\,GHz)& PKS-HART-CD-HO-MP-AT        & 0.60        & $0.40\pm0.06 $ & 2.61 & 1.64 & 0.41 & 8 \\
2008-06-09 (8.4\,GHz)& PKS-HART-CD-HO-MP-AT        & 1.06     & $0.63\pm0.09$ & 3.11 & 2.86 & 1.18 & $-123$\\
2008-11-27 (8.4\,GHz)& PKS-CD-HO-MP-AT-DSS43-TC-OH & 0.74  & $0.37\pm0.06 $ & 3.91 &  0.98 & 0.59 & 31 \\
2009-09-05 (8.4\,GHz)& PKS-CD-HO-MP-AT-DSS43-TC-OH & 0.76        & $0.45\pm0.07$ & 3.97 & 2.29 & 0.58 & 16 \\
2009-12-13 (8.4\,GHz)& PKS-CD-HO-MP-AT-TC          & 1.03        & $0.18\pm0.03 $ & 3.82 & 3.33 & 0.78 & 26 \\
2010-07-24 (8.4\,GHz)& PKS-CD-HO-MP-AT-TC          & 1.21       & $0.38\pm0.06 $ & 4.20 & 2.60 & 0.87 & 21\\
2011-04-01 (8.4\,GHz) & HART-CD-HO-MP-AT-DSS43-WW   & 0.63     & $0.31\pm0.05 $ & 5.10 & 2.31 & 0.51 &$-1$\\
\hline\hline
\textbf{uniform weighting} & & & & & & & \\

2008-11-27  (8.4\,GHz)&PKS-CD-HO-MP-AT-DSS43-TC-OH& 0.48   & $0.45\pm0.02$ & 3.2 &0.68      &0.43   &    33 \\
2008-11-29 (22.3\,GHz)& PKS-CD-HO-MP-AT-DSS43& 1.46 & $1.20\pm0.10$     & 3.3&1.55 &1.21   & $-75$          \\
\hline

\end{tabular}\\
\end{center}
{\footnotesize
$^{a}$ AT: Australia Telescope Compact Array, CD: Ceduna, HART:
Hartebeesthoek, HO: Hobart, MP: Mopra,  OH: GARS/O'Higgins, PKS:
Parkes, TC: TIGO, TI: DSS43 - NASA's Deep Space Network Tidbinbilla
(70\,m), WW: Warkworth; $^{b-g}$ image parameters: peak flux density, RMS noise, total
flux density, major and minor axis and position angle of the
restoring beam. We estimate a flux density uncertainty of $\sim$15\%.
}
\end{table*}

\subsection{High-resolution imaging}
The high-resolution images show a well collimated, straight jet with a small
opening angle of $\lesssim12^\circ$ \citep[compared to the rapid broadening
observed in M\,87,][]{Junor1999}. The counterjet is significantly
detected in all images.
The jet is resolved in unprecedented detail and into distinct
features. The emission is detected up to a maximum extent of
$\sim$70\,mas from the image phase center, i.e., TANAMI observations
probe the inner parsec of Cen~A with an angular resolution of less than
$\sim$0.01\,pc.
The VLBI core of the jet is identified as the brightest feature. Next
to the core, the second brightest, isolated jet component (at a
distance of
$\sim$3.5\,mas) is found to be not moving and stable in flux. It can be identified with
the stationary component discussed in
\citet{Tingay2001b}. Extragalactic jets often show stationary
components which can be explained as locally beamed emission or
standing shocks \citep{Lister2009a}. A possible interpretation of this
feature is a local pressure enhancement (like a jet nozzle) or a
cross-shock in the jet flow as observed in theoretical simulations of
over-pressured jets \citep{Mimica2009a}. Further (dual-frequency) VLBI
monitoring is required to investigate this in detail.

\subsection{The ``tuning fork''}
At a projected distance of $\sim$0.4\,pc from the jet core, the VLBI
images show a ``tuning-fork'' like structure, where the jet flow is
interrupted and widens up, but remains collimated downstream. The
brightness profile along the jet \citep[see Fig.~3 in][]{Mueller2014b}
shows a sharp gap.  The surface brightness locally decreases and no
positional change is observed, i.e., the feature remains stationary
over the monitoring period. In \citet{Mueller2014b}, we discuss this
feature and possible interpretations in detail. We conclude that while
it resembles a recollimation shock, decelerating and separating the
jet flow as seen in simulations by \citet{Perucho2007}, the overall
structure is difficult to reconcile with this scenario.  As a
different explanation we consider a standing disturbance, causing an
interaction with the jet. In particular, we discuss the penetration of
a gas cloud or massive star. Simulations by \citet{BoschRamon2012},
describing jet-star interactions in AGN, show similar resulting
bow-shock structures. Note that such interactions in Cen~A's jet have
already been considered by \citet{Hardcastle2003} to explain the
observed X-ray knots in the kpc-scale jet.  We find that a red giant
with a stellar wind of $v\approx100\,\mathrm{km\,s^{-1}}$ and a mass
loss rate of $10^{-8}\mathrm{M_\odot yr^{-1}}$ could create an
equilibrium condition, such that it can penetrate the jet flow without
disruption. Detailed estimations and calculations can be found in
\citet{Mueller2014b}.

In \citet{Mueller2015b}, we discuss the consequences of such an
interaction event. Theoretical simulations expect $\gamma$-rays from
the collision of a red giant or a massive object with the jet
plasma
\citep{Araudo2013,Khangulyan2013,BoschRamon2012,Barkov2010}, producing
variability on short time scales of hours to days. Cen~A is a bright
$\gamma$-ray source, but it shows no significant variability
\citep{Abdo2010_cenacore,3fgl}. We estimate that this persistent emission could
be partly produced by multiple jet-star-interaction events.

\subsection{Sub-parsec scale jet kinematics}
\begin{figure}
\includegraphics[width=\columnwidth]{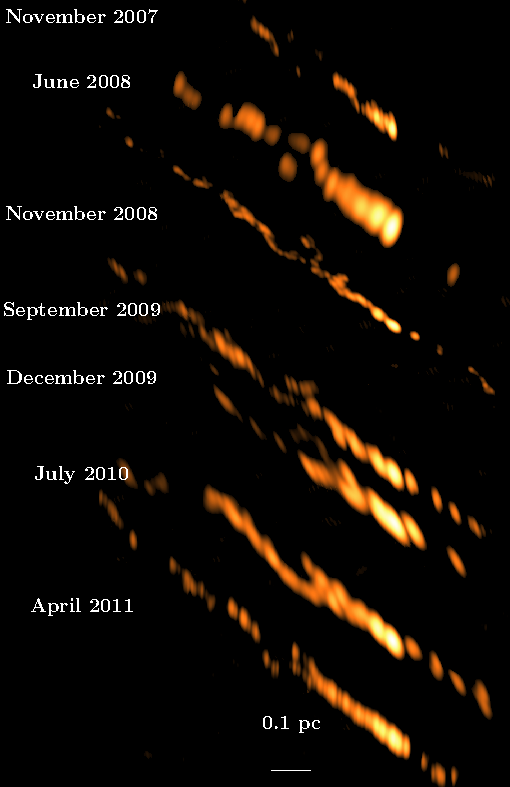}
\caption{Time evolution of the innermost parsec of Cen~A's jet. Shown
  are the individual VLBI images from TANAMI monitoring at 8\,GHz. The
  image parameters are given in Table~\ref{tablecena}. Note the
  different angular resolutions and beam shapes for each epoch due to
  the different configurations of the TANAMI VLBI array. See also the
  corresponding proper motion movie presented
  in \citet{Kadler2015}.
}
\label{fig:epochs}
\end{figure}

The time evolution of the pc-scale jet of Cen~A had been well
studied over a period of about twelve years by
\citet{Tingay1998b,Tingay2001b}. Tracking two moving components, they
determined an apparent speed of $\sim 2\,\mathrm{mas\,yr^{-1}}$
(corresponding to $\beta_\mathrm{app}=0.12$).  The 3.5\,years of
TANAMI monitoring allow us to study Cen~A's jet in more
detail. With about ten times better angular resolution we can
resolve the innermost parsec of the jet into multiple individual
components. The time evolution of the jet is presented in
Fig.~\ref{fig:epochs}. 

To parametrize and track the individual jet features, we
use the \texttt{modelfit} task in DIFMAP \citep{Shepherd1997} to fit
Gaussian emission model components to the self-calibrated visibility
data. 
Figure~6 in \citet{Mueller2014b} shows the kinematics, i.e., the time dependent seperation of each
component from the core. A linear regression fit gives the
proper motion for each component in mas/yr. 
As an illustration of the plasma
motion in the jet, we produced a movie using the individual VLBI
images interpolated in time (Fig.~\ref{fig:epochs}), as reported in \citet{Kadler2015}.  It
can be accessed at \url{http://www.aip.de/AN/movies}.

Besides the core and the stationary component next to it, we can
identify eight moving components showing a range of speeds from $\sim
1.8\,\mathrm{mas\,yr^{-1}}$ to $\sim 5\,\mathrm{mas\,yr^{-1}}$, with a
mean apparent speed of $\sim 2.98\,\mathrm{mas\,yr^{-1}}$.
Using the tapered visibility data we can show \citep{Mueller2014b} that this result is
consistent with the previous measurements by \citet{Tingay1998b,Tingay2001b}. 
Furthermore the comparison of the naturally weighted and tapered
images show that we can interpret this as an underlying jet flow with
faster substructure.

We further find that the outer components have higher apparent speeds
than the ones closer to the core, i.e., showing apparent
acceleration downstream. This result connects to the speed of $\sim 0.5\,c$ measured at
$\sim$100\,pc from the core by \citet{Hardcastle2003}, already suggesting
intrinsic acceleration from pc to kpc scales.
Furthermore, this region coincides with the optically thin part of the
jet \citep{Mueller2011a}. This correlation can naturally be explained
in the context of a spine-sheath structure of the jet: a faster, inner
spine, surrounded by a slower sheath. The larger speeds measured
in optically thin regions would correspond to the faster spine.

With the derived apparent speeds and the flux density
jet-to-counterjet ratios from the individual and stacked images, we can set
constraints on the intrinsic jet speed and jet inclination angle \citep[see also
Fig.~9 in][]{Mueller2014b}, following:
 \begin{equation}\label{eq:superluminal}
 v_\mathrm{app}=\frac{v \sin\theta}{1-\frac{v}{c}\cos\theta}
 \end{equation}
and
\begin{equation}\label{eq:R}
R=\frac{S_\mathrm{jet}}{S_\mathrm{counterjet}}=\left(\frac{1+\beta\cos\theta}{1-\beta\cos\theta}\right)^{2-\alpha}\quad,
 \end{equation}
with the apparent speed $v_\mathrm{app}$, the brightness ratio $R$ of
the approaching jet to the counterjet,
the inclination angle $\theta$, the intrinsic speed $\beta =
\frac{v}{c}$, and the spectral index $\alpha$.

TANAMI observations limit the intrinsic jet speed to
$\beta\sim0.24-0.37$ and the angle to the line of sight to
$\theta\sim12^\circ-45^\circ$.  The upper limits for the inclination
angle are consistent within the uncertainties with the results for the
pc-scale jet from \citet{Jones1996} and \citet{Tingay1998b}, but
our limits better match the results by \citet{Hardcastle2003} at larger
scales.  We find that a change of the angle to the line of sight
cannot explain the difference in speed. Therefore, the larger
speed at kpc scales can consistently be explained by intrinsic
acceleration.

\subsection{On the high-energy emission origin}
Cen~A is bright across the electromagnetic spectrum with a blazar-like
broadband SED of its core emission. The high-energy emission has been extensively
studied, but the origin and mechanism(s) are still debated. The radio
to $\gamma$-ray emission can be well
described by a single-zone synchrotron self-compton model
\citep{Abdo2010_cenacore}, but it fails to model the (non-simultaneous) TeV
emission. In order to better constrain theoretical models, we need to
disentangle the individual emission components and study them in detail.

In \citet{Mueller2011a}, we presented the first dual-frequency observation resulting
in the first sub-pc scale spectral index map for this source. The spectral index changes along the jet from flat/inverted
in the core, to steep further downstream. The central part of the
jet-counterjet region indicates free-free absorption as previously
discussed by \citet{Tingay2001a}.  The spectral index distribution along
the jet shows multiple bright, compact, and optically thick regions,
which can be interpreted as possible production sites of highly
energetic photons. As the origin of the $\gamma$-ray emission in
extragalactic jets is still debated, this result suggests that
multi-zone emission models need to be considered.

Furthermore the origin of the hard X-ray emission is still not clear. Several results
indicate the jet as the possible origin
\citep{Tingay1998b,Fukazawa2011,Beckmann2011a}. In \citet{Tingay1998b}
the coincidence of two ejection events with X-ray high-flux states are
discussed, suggesting a relation of jet and X-ray activity.  
Following this study, we use archival X-ray monitoring data by
\textsl{Swift}/BAT, \textsl{RXTE}/ASM, \textsl{RXTE}/PCA,
\textsl{CGRO}/BATSE, and the 90-350\,GHz lightcurve from SEST
\citep{Israel2008} to compare to the VLBI flux density variability at
8\,GHz from \citet{Tingay2001b} and TANAMI monitoring \citep[see
Fig.~10 in][]{Mueller2014b}.
The analysis of the jet
kinematics using TANAMI data revealed two new components ejected into
the jet during the 3.5\,year monitoring period. Their ejection time
can be estimated to be between 2007 and 2009. We find a partial
overlap of higher X-ray activity and these jet ejections, although
we cannot claim a common origin based on this result; more correlated
events need to be found. 

Recently, \citet{Fuerst2016} performed a detailed study of the
3--78\,keV emission using simultaneous observations by \textsl{NuSTAR}
and \textsl{XMM-Newton}. 
The sub-arcmin imaging with \textsl{NuSTAR} results in no evidence for
a jet detection above 10\,keV . 
The combined spectrum can be fitted either with an absorbed power-law
component ($\Gamma = 1.815\pm0.005$) or a single-temperature
Comptonization spectrum, and an additional flourescent Fe\,K$\alpha$
emission line. The physical origin of the observed hard X-ray spectrum
is discussed in detail, concluding that the emission can be explained
by synchrotron self-Compton emission from the inner jet or by an
advection-dominated accretion flow or a combination of both.
Further multi-epoch, multiwavelength observations will help us to
disentangle these scenarios. 

%%%%%%%%%%%%%%%%%%
\section{Conclusion and Outlook}
It has been discussed how combined multiwavelength and VLBI studies of
extragalactic jets can
shed light on the physics of these powerful objects. These observations
provide both, monitoring of source activity and changes in the
spectrum as well as
highly resolved images of the innermost regions, where the power is
thought to be released.

The monitoring of the TANAMI program is set up to address open
questions in jet physics. Two TANAMI sources have been studied in
great detail, namely \pmn and Cen~A. 
Both sources present ideal objects to study the high-energy emission
and formation of jets. 

\pmn is one of the brightest sources in the $\gamma$-ray sky, but
shows no major flaring activity. Its
unusual broadband properties question its classification as a blazar
and open room for an alternative interpretation.
Future observations can confirm the CSO classification. Only recently
\citet{Migliori2016} presented the first $\gamma$-ray detection of a
confirmed CSO (PKS\,1718-649). Since PMN\,1603-4904 has a hard $\gamma$-ray spectrum, it is a
likely candidate source for TeV instruments like H.E.S.S. or in future
CTA, and
therefore it could play an important role in investigating the
high-energy properties in misaligned sources.

The sub-pc scale imaging of Cen~A provides unprecedented insights
into the properties of the inner region of an AGN jet. We observe complex jet dynamics,
which, together with long-term light curves can help to constrain SED model
parameters. The overall jet structure can be well explained by a
spine-sheath configuration. Connecting our results for the
pc-scale jet and the observations at hundreds of parsecs requires
intrinsic acceleration between these scales.
Individual jet features can be studied in detail. The jet widening at
a distance of $\sim$0.4\,pc from the core could arise from a jet-star
interaction.

Thanks to the recent developments in VLBI at millimeter
wavelengths (mm-VLBI), we will be able to further study southern
extragalactic jets at even higher angular resolution. Future mm-VLBI
observations will include the Atacama Large Millimeter Array (ALMA) in
Chile, providing for the first time enough sensitivity and suitable
$(u,v)$-coverage to image sources below $-30^\circ$ declination at
millimeter wavelengths. 
In particular Cen~A presents an ideal target due to its proximity,
such that we can obtain insights into regions that are self-absorbed
at longer wavelengths and are located even closer to the jet base.

%%%%%%%%%%%%%%%%%%
\acknowledgements
I thank the committee of the German Astronomical Society for
awarding me with the Doctoral Thesis Prize 2015.

I thank Matthias Kadler, J\"orn Wilms, and Roopesh Ojha for their support, and all
collaborators of the TANAMI, \textsl{Fermi}/LAT and ANTARES teams for the
fruitful discussions and cooperation. Special thanks
to Robert Schulz and Felicia Krau\ss for proofreading the manusscript,
and to all
colleagues of the Remeis Observatory in Bamberg, at ECAP in Erlangen, and at the Chair of
Astronomy in W\"urzburg for the inspiring working environment.

I acknowledge the funding through a PhD fellowship from the
Studienstiftung des Deutschen Volkes and the support of the
Bundesministerium für Wirtschaft und Technologie (BMWi) through the 
Deutsches Zentrum für Luft- und Raumfahrt (DLR) grant 50OR1404.

This research has made use of a collection of ISIS scripts provided by
the Dr.~Karl Remeis observatory, Bamberg, Germany, at
\url{http://www.sternwarte.uni-erlangen.de/isis/}. The Long Baseline
Array and Australia Telescope Compact Array are part of the Australia
Telescope National Facility, which is funded by the Commonwealth of
Australia for operation as a National Facility managed by CSIRO.

\end{document}